\documentclass[]{elsarticle}

\usepackage{natbib}
\usepackage[rgb]{xcolor}
\journal{CAP}

\begin{document}
\begin{frontmatter}
\title{Comment on "Evidence for absence of metallic surface states in BiO$_2$-terminated BaBiO$_3$ thin films"}
\author[CAC,CONICET]{D. Rubi}
\author[CAC,CONICET]{A. M. Llois}
\author[CAC,CONICET]{V. L. Vildosola}
\address[CAC]{Departamento de Materia Condensada , GIyA, CNEA and Instituto de Nanociencia y Tecnolog\'{\i}a, 1650 San Mart\'{\i}n, Buenos Aires, Argentina}
\address[CONICET]{Consejo Nacional de Investigaciones Cient\'{\i}ficas y T\'ecnicas (CONICET), Argentina}

\begin{abstract}
In a recent work by Ji Seop Oh \textit{et al.}, BaBiO$_3$(001) thin films were grown on SrTiO$_3$ by Pulsed Laser Deposition. It was argued that the films are BiO$_2$-terminated from the modelling of 
angle-resolved photoemission spectroscopy experiments. The authors claim, in opposition to previous theoretical predictions, that there are no metallic surface states on their films. 
In this short comment we question that the authors have enough evidence to make such a claim, as we consider that the large mismatch between SrTiO$_3$ and BaBiO$_3$ and the lack of control of their 
fabrication process with reflection high energy electron diffraction could unlikely give high quality films with a single BiO$_2$-termination, which is one of the requisites for the stabilization 
of these surface metallic states.
\end{abstract}


\end{frontmatter}

Ji Seop Oh \textit{et al.} have recently studied the electronic structure of (001) oriented epitaxial BaBiO$_3$
(BBO) thin films grown on SrTiO$_3$ (STO) single crystals \cite{OH2018658}. The films were grown by means of the Pulsed Laser Deposition (PLD) technique. 
They claim, based on the modelling of in-situ angle-resolved 
photoemission spectroscopy (ARPES), that their films are BiO$_2$-terminated. A similar claim was previously done by Gozar \textit{et al.} \cite{PhysRevB.75.201402}. According to the theoretical work by Vildosola \textit{et al.} \cite{PhysRevLett.110.206805}, a (001) BiO$_2$-termination should trigger the appearance of a metallic 2D electron gas (2DEG) at the surface. In Vildosola's \textit{et al.} work, ab-initio calculations show that the two outermost layers of the BiO$_2$-terminated surface turn cubic-like and metallic, while the inner ones remain in the insulating monoclinic state that the system presents in its bulk form. On the other hand, the surface metallization is not predicted for the BaO-termination. Ji Seop Oh \textit{et al.} find, instead, a surface band structure inconsistent with a metallic state, and claim that the aforementioned theoretical work is inaccurate \cite{OH2018658}.

We consider that the claim of Ji Seop Oh \textit{et al.} (and previously by Gozar \textit{et al.} \cite{PhysRevB.75.201402} ) about the single BiO$_2$ termination of their films can be seriously put into question due to several reasons that we describe below. 

First, the lattice mismatch between BBO and STO ($\sim$12\%) is unusually large to allow high quality, single terminated, epitaxial films of BBO. 
Indeed, it has been shown that the structure of the BBO films on STO is similar to the bulk material \cite{PhysRevB.75.201402}, indicating the presence of a strong strain relaxation mechanism during the first growth states. This mechanism has been recently unveiled by Zapf \textit{et al.} \cite{zapf2018}, who have shown by high resolution transmission electron microscopy that strain relaxation occurs at the BBO/STO interface through the formation of an ultrathin layer of a different phase that accommodates the large misfit between both structures, allowing the BBO film formed on top of that layer to be unstrained. 
Importantly, the BBO films develop crystallographic domains, with typical sizes of ~10-40nm, and antiphase boundaries. In these extended defects, a lattice shift of half unit cell is 
found between adjacent grains in the direction perpendicular to the substrate. This indicates that a single atomic termination for BBO films grown on STO is hardly achievable. To circumvent this problem, 
substrates with lower mismatch such as LaLuO$_3$ should be used\cite{PhysRevMaterials.2.041801}, or buffers with intermediate cell parameters should be sandwiched between STO and 
BBO, in order to allow an efficient strain accommodation, avoiding the formation of extended defects. The latter strategy was followed in Ref. \citenum{Lee-2018}.

Second, the standard procedure to obtain a desired atomic termination for an ABO$_3$ perovskite grown by PLD on TiO$_2$-terminated STO is by using reflection high energy electron diffraction (RHEED) assisted 
deposition. In particular, the growth conditions should be optimized in order to get a layer-by-layer growth mechanism, and the RHEED intensity oscillations allows to control the film termination 
(either BO$_2$ or AO). In their paper, Ji Seop Oh \textit{et al.} do not report a growth process assisted by RHEED, so that we consider that it is highly unlikely to obtain a BBO single termination without 
this high control of the deposition process. In addition, they did not complement their studies with a BaO-terminated film, as one would expect. Based on these considerations, we think that the modelling of 
the ARPES data is not enough evidence to claim that the fabricated BBO films are BiO$_2$-terminated and that the measured surface band structure by Ji Seop Oh \textit{et al.} does not correspond to the single BiO$_2$-terminated (001) surface.

Finally, we would like to criticize the authors claim saying that the absence of a metallic surface state
could be attributed to the oversimplified theoretical calculation, which fails to give an accurate estimation of the band gap in BBO. 
It is true, as the authors say, that the theoretical value of the band gap depends strongly on the approximation done in the exchange-correlation functional. 
BBO is not an exception of this rule. This effect was studied in detailed in Ref. \citenum{PhysRevB.81.085213} and confirmed by some of us in the Supplementary Material of Ref. \citenum{PhysRevLett.110.206805}. 
In addition, the authors claim that the theoretical gap values of Ref. \citenum{PhysRevLett.110.206805} were much smaller than the lower bound of 1.9 eV they obtained from the ARPES spectra. 
Here, the authors should distinguish between the direct optical gap (E$_g$) that, for bulk BBO, is around 2 eV, and the smaller indirect gap, E$_i$. While the experimental value of E$_g$ is quite well accounted by several functionals used in the above mentioned theoretical works (regarding bulk BBO), there is some dispersion in the reported experimental values of E$_i$, ranging from 0.2 to 1.1 eV. The functional used both in Ref. \citenum{PhysRevLett.110.206805} and Ref. \citenum{PhysRevB.81.085213} give values of E$_i$ that lie well within that range except for the simplest local and semilocal functionals as LDA\cite{LDA} or
 GGA\cite{GGA}. In any case, we want to remark that in Ref. \citenum{PhysRevLett.110.206805} it was demonstrated that the predicted appearance of metallic states at the BiO$_2$-terminated surface is independent of the value of the band gap and it is robust against the functional used in the DFT calculation.

In summary, we consider that the conclusions of the work by Ji Seop Oh \textit{et al.} are inaccurate, as the chosen STO substrate is not the appropriate to obtain high quality single terminated BBO films due to the large mismatch between both structures. The experimental validation of the mechanism theoretically proposed by Vildosola  \textit{et al.} waits for the band structure characterization of highly controlled BiO$_2$-terminated BBO films grown on lower mismatch substrates.
                         
The authors acknowledge finantial support from ANPCyT. In particular, D. Rubi and A. M. Llois are supported by  PICT-2016-0867. D. Rubi also thanks support from PICT 2017-1836 while A. M. Llois from 
PICT-2014-1555. Finally, V. Vildosola thanks  PICT 2015 0869 and  PICTE 2014 134. 

\bibliography{bbo}
\bibliographystyle{elsarticle-num} 

\end{document}